\documentclass[amsmath,amssymb]{revtex4}
\usepackage{graphicx}
\usepackage{dcolumn}
\usepackage{bm}

\begin{document}
\title[]
{A note on wormholes in slightly modified
   gravitational theories}
\author{Peter K.\,F. Kuhfittig}
\address{Department of Mathematics\\
Milwaukee School of Engineering\\
Milwaukee, Wisconsin 53202-3109}

\begin{abstract}\noindent
Wormholes that meet the flare-out condition violate the
null energy condition in classical general relativity.
The purpose of this note is to show that even a slight
modification of the gravitational theory could, under
certain conditions, avoid this violation for the matter 
threading the wormhole.  The first part discusses some 
general criteria based on the field equations, while 
the second part assumes a specific equation of state 
describing normal matter, together with a particular 
type of shape function.  The analysis is confined to 
wormholes with zero tidal forces.

\phantom{a}

\end{abstract}

\maketitle

\section{Introduction}\noindent
Interest in modified theories of gravity has increased
greatly since the discovery that our Universe is undergoing
an accelerated expansion.  In particular, $f(R)$ modified
gravity replaces the Ricci scalar $R$ in the
Einstein-Hilbert action
\begin{equation}\label{action1}
  S_{\text{EH}}=\int\sqrt{-g}\,R\,d^4x
\end{equation}
by a nonlinear function $f(R)$:
\begin{equation}\label{action2}
   S_{f(R)}=\int\sqrt{-g}\,f(R)\,d^4x.
\end{equation}
(For a review, see Refs. \cite{SF08, NO07, fL08}.)
Wormhole geometries in $f(R)$ modified gravitational
theories are discussed in Ref. \cite{LO09}.  Ref.
\cite{J} assumes a noncommutative-geometry background
in constructing wormhole geometries in $f(R)$ gravity.

It is well known that wormholes in classical general
relativity (GR) require a violation of the null energy
condition, usually calling for the use of ``exotic
matter"  \cite{MT88}.  Such matter must be confined
to a very narrow band around the throat \cite{FR96,
Kuhf}.  Dealing with a very small region suggests
that a small modification of the gravitational theory
may take the place of exotic matter, analogous to the
way that the smearing effect in noncommutative
geometry can replace such exotic matter \cite{pK13}.
So in studying the effect of the slightly modified
gravity, we concentrate mainly on the vicinity of
the throat.

\section{Wormhole geometries in slightly modified $f(R)$
     gravity}\noindent
To describe a spherically symmetric wormhole spacetime,
we take the metric to be \cite{MT88}
\begin{equation}
    ds^2=-e^{\Phi(r)}dt^2+\frac{dr^2}{1-b(r)/r}
    +r^2(d\theta^2+\text{sin}^2\theta\,d\phi^2).
\end{equation}
Here we recall that $b=b(r)$ is called the
\emph{shape function} and $\Phi=\Phi(r)$ the
\emph{redshift function}.  For the shape function
we must have $b(r_0)=r_0$, where $r=r_0$ is the
radius of the \emph{throat} of the wormhole.  In
addition, $b'(r_0)<1$ and $b(r)<r$ to satisfy the
\emph{flare-out condition} \cite{MT88}. These
restrictions result in the violation of the null
energy condition in classical general relativity,
especially in the vicinity of the throat.

Regarding the redshift function, we normally require
that $\Phi(r)$ remain finite to prevent an event
horizon.  In the present study involving $f(R)$
gravity, we need to assume that $\Phi(r)\equiv
\text{constant}$, so that $\Phi'\equiv 0$.
Otherwise, according to Lobo \cite{LO09}, the
analysis becomes intractable.

Our next task is to define what is meant by
slightly modified gravity.  To this end, we list
the gravitational field equations in the form used
by Lobo \cite{LO09}:
\begin{equation}\label{FE1}
   \rho(r)=F(r)\frac{b'(r)}{r^2},
\end{equation}
\begin{equation}\label{FE2}
  p_r(r)=-F(r)\frac{b(r)}{r^3}+F'(r)
  \frac{rb'(r)-b(r)}{2r^2}-F''(r)
     \left[1-\frac{b(r)}{r}\right].
\end{equation}
and
\begin{equation}\label{FE3}
  p_t(r)=-\frac{F'(r)}{r}\left[1-\frac{b(r)}{r}
  \right]+\frac{F(r)}{2r^3}[b(r)-rb'(r)],
\end{equation}
where $F=\frac{df}{dR}$.  The curvature scalar $R$ is
given by
\begin{equation}\label{Ricci}
   R(r)=\frac{2b'(r)}{r^2}.
\end{equation}
While it is possible in principle to obtain $f(R)$
from $F(r)$, our goal is more modest: how to define
\emph{slightly modified $f(R)$ gravity}.  To this
end, we observe that the above field equations
reduce to the Einstein equations for $\Phi'\equiv 0$
whenever $F\equiv 1$.  Consequently, comparing Eqs.
(\ref{FE1}) and (\ref{Ricci}), a slight change in
$F$ results in a slight change in $R$, which,
referring to Eqs. (\ref{action1}) and
(\ref{action2}), characterizes $f(R)$ modified
gravity.  So we may quantify the notion of slightly
modified gravity by assuming that $F(r)$  remains
close to unity and relatively $``\text{flat},"$
i.e., both $F'(r)$ and $F''(r)$ remain relatively
small in absolute value.  To discuss wormholes, we
need the additional assumption that $F'(r_0)$ is
negative. (We will see in the next section that
this condition is actually met when $F(r)$ is
computed from a known shape function.) $F''(r)$
will be discussed later. Observe that in
Eq. (\ref{FE1}), $F$ behaves like a
dimensionless scale factor.

Suppose the shape function meets the flare-out
condition $b'(r_0)<1$.  By continuity, $b(r)<r$
in the immediate vicinity of the throat.  Our goal
is to show that in this region, we may have
$\rho+p_r \ge 0$, as well as $\rho+p_t\ge 0$,
thereby satisfying the null energy condition.
From Eqs. (\ref{FE1}) and (\ref{FE2})
\begin{equation}\label{NE1}
   \rho+p_r=F\frac{b'}{r^2}-F\frac{b}{r^3}+F'
   \frac{rb'-b}{2r^2}-F''\left(1-\frac{b}{r}
   \right)=(rb'-b)\left(\frac{F}{r^3}
   +\frac{F'}{2r^2}\right)-F''\left(1-\frac{b}{r}
   \right)\ge 0.
\end{equation}
Since $1-b/r$ is close to zero near the throat,
let us disregard the last term for now.  The
flare-out condition implies that
$rb'(r)-b(r)<0$; so we must have
\[
   \frac{F}{r^3}+\frac{F'}{2r^2}\le 0.
\]
Given that $F(r)$ is very close to unity near
the throat and that $F'(r)<0$ and relatively
small in absolute value, $F/r_0^3+F'/2r_0^2$
can only be negative if $r_0$ is sufficiently
large.  Consider a simple example.  In the
vicinity of the throat, suppose that
\begin{equation*}
   F=2-e^{a(r-r_0)};\quad \text{then}\quad
   F'=-e^{a(r-r_0)}a\quad \text{and}\quad
   F''=-e^{a(r-r_0)}a^2,
\end{equation*}
where $a$ is a small positive constant.  At
$r=r_0$, $F(r_0)=1$, $F'(r_0)=-a$, and
$F''(r_0)=-a^2$.  Substituting in Eq. (\ref{NE1}),
we get
\[
  (rb'-b)\frac{2-ar_0}{2r_0^3}+a^2
      \left(1-\frac{b}{r}\right)\ge 0,
\]
provided that $2-ar_0\le 0$.  We conclude that
\begin{equation}\label{large1}
   r_0\ge \frac{2}{a}.
\end{equation}
For example, if $a=0.001\, \text{m}^{-1}$, then
$r_0\ge 2\,\text{km}$.  The general conclusion is
that
\begin{equation}\label{large2}
   r_0\ge \frac{2}{-F'(r_0)},\,\,\,\text{where}
   \,\,\,F'(r_0)<0.
\end{equation}
Observe that $F''(r_0)$ must either be negative or
negligibly small.  (In the above example, it is
actually both.)  Finally, by Eq. (\ref{FE3}), we
also have $\rho+p_t\ge 0$.

The closer $F'(r_0)$ is to zero, the larger $r_0$ has
to be to meet the condition $\rho +p_r\ge 0$.
Returning to Einstein gravity, if $F'(r_0)
\rightarrow 0$, then $r_0\rightarrow\infty$, and we
do not get a wormhole.  In this case, then, the
existence of a wormhole requires that $\rho+p_r<0$,
the usual violation of the null energy condition in
GR.  As we have seen, we also get $\rho+p_r<0$
whenever $F'(r_0)\ge 0$.

\section{A known shape function}\noindent
The above analysis assumes a small change in the Ricci
scalar, induced by a small change in $F$.  An
alternative approach is to assume a certain equation
of state and a known shape function and then
determine $F$ in the vicinity of the throat.  To
make the analysis tractable, $b(r)$ must be
relatively simple, yet typical enough to yield a
reasonably general conclusion.

The equation of state to be used for now assumes
normal matter:
\begin{equation}\label{EoS}
   p_r=\omega\rho,\quad 0<\omega <1.
\end{equation}
For the shape function we use the form
\begin{equation}\label{shape}
  b(r)=r_0^{1-\alpha}r^{\alpha},\quad 0<\alpha <1.
\end{equation}
Observe that $b(r_0)=r_0$ and $b'(r_0)=\alpha<1$,
so that the flare-out condition has been met.

As before, we want $F(r_0)\approx 1$, but for
computational convenience, we assume that
$F(r_0)=1$.  From Eqs. (\ref{FE1}) and (\ref{FE2}),
\begin{equation}\label{NE2}
  \omega F\frac{r_0^{1-\alpha}\alpha r^{\alpha-1}}{r^2}
  =-F\frac{r_0^{1-\alpha}r^{\alpha}}{r^3}
  +F'\frac{rr_0^{1-\alpha}\alpha r^{\alpha-1}-
  r_0^{1-\alpha}r^{\alpha}}{2r^2}-F''
    \left(1-\frac{r_0^{1-\alpha}r^{\alpha}}{r}\right).
\end{equation}
According to Lobo \cite{LO09}, the existence of $F''$
makes it virtually impossible in most cases to get an
exact solution.  While the last term is once again
zero at the throat, we still need to be concerned with
the vicinity of the throat. To this end, we need to
assume that $\alpha$ is close to unity, so that the
last term is close to zero.  What remains is easy
enough to solve.  After simplifying, we obtain the
linear differential equation
\[
   F'(r)+\frac{2(1+\alpha\omega)}{1-\alpha}\frac{1}{r}
   F(r)=0,
\]
where $r$ is near the throat.  The solution is
\begin{equation}\label{solution1}
   F(r)=\left(\frac{r}{r_0}\right)^
   {-2(1+\alpha\omega)/(1-\alpha)},
\end{equation}
which satisfies the condition $F(r_0)=1$ and so
$F(r)\approx 1$ in the vicinity of the throat.
Having satisfied the normal-matter equation,
Eq. (\ref{EoS}), in the vicinity of the throat,
as well as the flare-out condition, we conclude
that the wormhole is sustained due to the
modified gravity.  (See Ref. \cite{LO09} for 
further details.)

Even though the last term in Eq. (\ref{NE2}) was
neglected, we are still dealing with a second-order
equation; so we need to consider $F'(r)$ in the
vicinity of the throat:

\begin{equation}\label{derivative1}
  \left. F'(r_0)=\frac{-2(1+\alpha\omega)}{1-\alpha}
   \left(\frac{r}{r_0}\right)^{-2(1+\alpha\omega)
   /(1-\alpha)-1}\frac{1}{r_0}\right|_{r=r_0}=
   \frac{-2(1+\alpha\omega)}{1-\alpha}\frac{1}
   {r_0}<0,
\end{equation}
i.e., $F'(r_0)<0$, as in the previous section, but
we still want $|F'(r_0)|$ to be relatively small
in order to remain close to Einstein gravity.  So
once again, $r_0$ has to be sufficiently large:
\begin{equation}\label{large3}
   r_0\ge \frac{2}{-F'(r_0)}\frac{1+\alpha\omega}
       {1-\alpha}.
\end{equation}
For example, if $\alpha=0.99$, $F'(r_0)=-0.3$, and
$\omega=0.5$, we obtain $r_0\ge 1\,\text{km}$.  (It
is readily checked that if $|F'(r_0)|$ is small,
then so is $|F''(r_0)|$.)

The parameter $\omega$ was chosen to describe normal
matter.  However, solution (\ref{solution1}) is valid
for any $\omega$.  Thus if $\omega=-1$, which is
equivalent to assuming Einstein's cosmological
constant \cite{mC01}, inequality (\ref{large3}) reduces
to inequality (\ref{large2}).  If $\omega<-1$, we are
dealing with phantom energy, which is known to support
wormholes in classical GR \cite{pK09, sS05, fL05, oZ05}.
In the present situation, we still need to consider
$\alpha$.  So if we assume that
\[
     -\frac{1}{\omega}<\alpha<1,
\]
then $(1+\alpha\omega)/(1-\alpha)$ becomes negative
and condition (\ref{large3}) is automatically
satisfied for all $r_0$.  We conclude that phantom
energy can also support wormholes in our slightly
modified gravitational theory.

\emph{A remark concerning $F''$:} Retaining $F''$ is
possible if $b(r)$ is sufficiently simple.  Suppose
$b(r)=ar$, $a>0$.  Then $b(r_0)=ar_0 \approx r_0$,
provided that $r_0$ is large compared to $a$.  Then
Eq. (\ref{NE1}) becomes
\[
   F''+\frac{1}{r^2}\frac{a(1+\omega)}{1-a}F=0,
     \quad F(r_0)=1,\quad 0<\omega<1.
\]
The solution is
\begin{equation}
   F(r)=\left(\frac{r}{r_0}\right)
   ^{\frac{1}{2}[1+\sqrt{1-4a(1+\omega)/(1-a)}]}.
\end{equation}
Given that $a$ is a small constant,
$F'(r_0)\approx 1/r_0$ and $F''(r_0)\approx -1/r_0^2$.
Since $r_0$ is assumed to be large, $F'(r)$ and
$|F''(r)|$ are small in the vicinity of the throat.

\section{Conclusion}\noindent
Wormholes that meet the flare-out condition violate the
null energy condition in classical general relativity.
It is shown in the first part of this note that a
slight modification of the field equations, and hence
of the gravitational theory, could avoid this violation
for the matter threading the wormhole. The modification 
calls for a change in $F(r)$, which induces a change 
in $R$ in the Einstein-Hilbert action: for every $r$, 
$F(r)$ remains close to unity, while $F'(r)$ and 
$F''(r)$ are relatively small in absolute
value.  The null energy condition is met provided that
$r_0$, the radius of the throat, is sufficiently large
and that $F'(r_0)<0$; $F''(r_0)$ must be either
negative or negligibly small.

The second part assumes the equation of state
$p_r=\omega\rho$, $0<\omega <1$, thereby representing
normal matter, as well as a shape function of the form
$b(r)=r_0^{1-\alpha}r^{\alpha}$, $0<\alpha <1$; hence
$b'(r_0)=\alpha <1$.  The equation of state yields the
solution $F(r)=(r/r_0)^{-2(1+\alpha\omega)/(1-\alpha)}$
in the vicinity of the throat, where $\alpha$ is close
to unity and $r_0$ is sufficiently large.  Since the
wormhole consists of ordinary matter, its survival
must be attributed to the modified gravity \cite{LO09}.
Moreover, a small modification would likely be
consistent with observation.

The assumption that $\omega$ is positive is not
actually necessary: in particular, phantom dark energy
can support a wormhole in slightly modified
gravitational theory, as well as in GR.  Zero tidal
forces are assumed throughout.

\end{document}